\documentclass[prc,twocolumn,showpacs,amsmath,amssymb,superscriptaddress,floatfix]{revtex4}
\usepackage{graphicx}
\usepackage{dcolumn}
\usepackage{bm}
\usepackage[dvips]{epsfig}
\usepackage{subfigure}

\newcommand{\reaction}{\mbox{$pp\to ppK^+K^-$}}
\newcommand{\Kmp}{\mbox{$K^-\!p$\ }}
\begin{document}
\title{Kaon Pair Production in Proton--Proton Collisions}
\author{Y.\,Maeda}\email[E-mail: ]{ymaeda@rcnp.osaka-u.ac.jp}
\affiliation{Research Center for Nuclear Physics, Osaka
University, Ibaraki, Osaka 567-0047, Japan}
\author{M.\,Hartmann}\email[E-mail: ]{M.Hartmann@fz-juelich.de}%
\affiliation{Institut f\"ur Kernphysik, Forschungszentrum
J\"ulich, 52425 J\"ulich, Germany}
\author{I.\,Keshelashvili}
\affiliation{Institut f\"ur Kernphysik, Forschungszentrum
J\"ulich, 52425 J\"ulich, Germany} \affiliation{High Energy
Physics Institute, Tbilisi State University, 0186 Tbilisi,
Georgia}
\author{S.\,Barsov}
\affiliation{High Energy Physics Department, Petersburg Nuclear
Physics Institute, 188350 Gatchina, Russia}
\author{M.\,B\"uscher}
\affiliation{Institut f\"ur Kernphysik, Forschungszentrum
J\"ulich, 52425 J\"ulich, Germany}
\author{M.\,Drochner}
 \affiliation{Zentralinstitut f\"ur Elektronik,
Forschungszentrum J\"ulich, 52425 J\"ulich, Germany}
\author{A.\,Dzyuba}
\affiliation{Institut f\"ur Kernphysik, Forschungszentrum
J\"ulich, 52425 J\"ulich, Germany} \affiliation{High Energy
Physics Department, Petersburg Nuclear Physics Institute, 188350
Gatchina, Russia}
\author{V.\,Hejny}
\affiliation{Institut f\"ur Kernphysik, Forschungszentrum
J\"ulich, 52425 J\"ulich, Germany}
\author{A.\,Kacharava}
\affiliation{High Energy Physics Institute, Tbilisi State
University, 0186 Tbilisi, Georgia} \affiliation{Physikalisches
Institut II, Universit\"at
  Erlangen--N\"urnberg, 91058 Erlangen, Germany}
\author{V.\,Kleber}
\affiliation{Physikalisches Institut, Universit\"at Bonn, 53115
Bonn, Germany}
\author{H.R.\,Koch}
\affiliation{Institut f\"ur Kernphysik, Forschungszentrum
J\"ulich, 52425 J\"ulich, Germany}
\author{V.\,Koptev}
\affiliation{High Energy Physics Department, Petersburg Nuclear
Physics Institute, 188350 Gatchina, Russia}
\author{P.\,Kulessa}
\affiliation{H.~Niewodnicza\'{n}ski Institute of Nuclear Physics
PAN, 31342 Krak\'{o}w, Poland}
\author{B.\,Lorentz}
\affiliation{Institut f\"ur Kernphysik, Forschungszentrum
  J\"ulich, 52425 J\"ulich, Germany}
\author{T.\,Mersmann}
\affiliation{Institut f\"ur Kernphysik, Universit\"at M\"unster,
48149 M\"unster, Germany}
\author{S.\,Mikirtytchiants}
\affiliation{Institut f\"ur Kernphysik, Forschungszentrum
J\"ulich, 52425 J\"ulich, Germany} \affiliation{High Energy
Physics Department, Petersburg Nuclear Physics Institute, 188350
Gatchina, Russia}
\author{A.\,Mussgiller}
\affiliation{Physikalisches Institut II, Universit\"at
  Erlangen--N\"urnberg, 91058 Erlangen, Germany}
\author{M.\,Nekipelov}
\affiliation{Institut f\"ur Kernphysik, Forschungszentrum
  J\"ulich, 52425 J\"ulich, Germany}
\affiliation{High Energy Physics Department, Petersburg Nuclear
Physics Institute, 188350 Gatchina, Russia}
\author{H.\,Ohm}
\affiliation{Institut f\"ur Kernphysik, Forschungszentrum
  J\"ulich, 52425 J\"ulich, Germany}
\author{D.\,Prasuhn}
\affiliation{Institut f\"ur Kernphysik, Forschungszentrum
  J\"ulich, 52425 J\"ulich, Germany}
\author{R.\,Schleichert}
\affiliation{Institut f\"ur Kernphysik, Forschungszentrum
  J\"ulich, 52425 J\"ulich, Germany}
\author{H.J.\,Stein}
\affiliation{Institut f\"ur Kernphysik, Forschungszentrum
  J\"ulich, 52425 J\"ulich, Germany}
\author{H.\,Str\"oher}
\affiliation{Institut f\"ur Kernphysik, Forschungszentrum
  J\"ulich, 52425 J\"ulich, Germany}
\author{Yu.\,Valdau}
\affiliation{Institut f\"ur Kernphysik, Forschungszentrum
  J\"ulich, 52425 J\"ulich, Germany}
\affiliation{High Energy Physics Department, Petersburg Nuclear
Physics Institute, 188350 Gatchina, Russia}
\author{C.\,Wilkin}
\affiliation{Physics and Astronomy Department, UCL, Gower Street,
London WC1E 6BT, UK}
\author{P.\,W\"ustner}
\affiliation{Zentralinstitut f\"ur Elektronik, Forschungszentrum
J\"ulich, 52425 J\"ulich, Germany}
%
\date{\today}

\begin{abstract}
The differential and total cross sections for kaon pair production
in the \reaction\ reaction have been measured at three beam
energies of 2.65, 2.70, and 2.83\,GeV using the ANKE magnetic
spectrometer at the COSY--J\"ulich accelerator. These
near--threshold data are separated into pairs arising from the
decay of the $\phi$--meson and the remainder. For the non--$\phi$
selection, the ratio of the differential cross sections in terms
of the \Kmp and $K^+p$ invariant masses is strongly peaked towards
low masses. This effect can be described quantitatively by using a
simple \emph{ansatz} for the \Kmp final state interaction, where
it is seen that the data are sensitive to the magnitude of an
effective \Kmp scattering length. When allowance is made for a
small number of $\phi$ events where the $K^-$ rescatters from the
proton, the $\phi$ region is equally well described at all three
energies. A very similar phenomenon is discovered in the ratio of
the cross sections as functions of the $K^-\!pp$ and $K^+pp$
invariant masses and the identical final state interaction model
is also very successful here. The world data on the energy
dependence of the non--$\phi$ total cross section is also
reproduced, except possibly for the results closest to threshold.
\end{abstract}
%
%
\pacs{13.60.Le, 
      14.40.Aq, 
      25.40.Ep  
}
\maketitle
%
%
\section{Introduction}
\label{sec:introduction}%

There are several mechanisms that can lead to the production of
kaon--antikaon pairs in nucleon--nucleon collisions near
threshold. These can be divided mainly into two general classes
involving (a) the production of a non--strange meson that
subsequently decays into $K\bar{K}$, and (b) the associated
production of $KY^*$, where the $\bar{K}$ is formed through the
decay of the hyperon $Y^*$. The most prominent meson that is found
almost exclusively in the $K^+K^-/K^0\bar{K}^0$ channels is, of
course the $\phi$, and there have been measurements of the $pp\to
pp\phi\to ppK^+K^-$ reaction from DISTO~\cite{Balestra} and
COSY--ANKE~\cite{Har06}, as well as of the $pn\to d\phi\to
dK^+K^-$ reaction~\cite{Mae06}. These experiments show strength
also away from the region of the $\phi$ peak that could arise from
other mesons decaying into $K^+K^-$.

The lightest of the non--strange scalar mesons are the $a_0(I=1)$
and $f_0(I=0)$, which decay mainly into $\eta\pi$ and $\pi\pi$,
respectively. Both have masses around 980\,MeV/c$^2$ and widths of
the order of 50--100\,MeV/c$^2$~\cite{PDG06}. Since the central
mass values fall almost exactly at the $K\bar{K}$ threshold, the
strong coupling to this channel distorts significantly the upper
parts of the mass spectra~\cite{PDG04}. It has even been suggested
that both resonances might be mainly molecular in
nature~\cite{Isgur}. Since the $K^+K^-$ system is a mixture of
isospin $I=0$ and $I=1$, one can consider whether the non--$\phi$
events observed in \reaction\ might be linked to the production
and decay of the $a_0$ and/or the $f_0$ mesons. The $I=1$ channel
is isolated cleanly by looking at the $pp\to dK^+\bar{K}^0$
reaction~\cite{Vera}.

On the other hand, we can also expect kaon pairs to be created in
the $pp\to K^+pY^*$ reaction, where the hyperon decays through
$Y^*\to K^-p$. There are, of course, several excited hyperons that
could contribute to such a process. Of particular interest for low
energy production are the $\Sigma(1385)$ and the $\Lambda(1405)$.
Though nominally lying below the sum of the $K^-$ and $p$ masses,
their large widths ($\sim 50\,$MeV/c$^2$) ensure that they overlap
the \Kmp threshold~\cite{PDG06}. Results have recently been
presented on the production of these states in the $pp\to
K^+p\Sigma^0(1385)$ and $pp\to K^+p\Lambda(1405)$ reactions, where
the hyperons were detected in the $\Lambda\pi^0$ and
$\Sigma^0\pi^0$ channels, respectively~\cite{Iza}.

In addition to being involved in a direct production mechanism,
the $\Lambda(1405)$ and the $\Sigma(1385)$ might also lead to a
significant \Kmp final state interaction (\emph{fsi}) in the
\reaction\ reaction.  Evidence that the \Kmp\ \emph{fsi} is indeed
important is to be found in the measurement of the \reaction\
cross section at energies below the $\phi$
threshold~\cite{Winter}. It was shown that the invariant masses of
the \Kmp final system were on average lower than those of the
$K^+p$ pairs and it was suggested that this is due to the strong
attraction in the \Kmp\ system.

It is therefore an interesting but open question whether the
\reaction\ data from the non--$\phi$ region should be viewed, in
first approximation, as scalar resonance production \emph{via}
$pp\to pp\,a_0/f_0$ or hyperon production through $pp\to
K^+p\Lambda(1405)/\Sigma^0(1385)$. It is one aim of the present
paper to study this competition by using the data obtained when
investigating $\phi$--meson production in proton--proton
collisions at ANKE~\cite{Har06}.

A program to investigate the production of kaon pairs in
nucleon--nucleon collisions has been initiated at the COSY storage
ring of the Forschungszentrum J\"ulich. Although the methods of
the measurement and analysis of the \reaction\ reaction have
already been given in Ref.~\cite{Har06}, for the sake of clarity
the main points are repeated in Sect~\ref{Experiment}. Due to the
finite width of the $\phi$, distinguishing between events
corresponding to the production of this meson and the rest
requires a detailed modelling of the $K^+K^-$ spectrum, as
described in Sect.~\ref{KKbar}. When more than one pair of
particles interact strongly in the final state, there is no
reliable prescription to evaluate a corresponding enhancement
factor. Our approach to this problem in terms of an effective \Kmp
scattering length $a$ is described in Sect.~\ref{fsia}, with a
heuristic justification being given in Appendix~\ref{app_a}.

The relative distributions of the \Kmp\ and $K^+p$ invariant
masses are shown and discussed in Sect.~\ref{Ratio}. The ratio of
the differential cross sections as functions of the \Kmp and
$K^+p$ invariant masses indicates that the \Kmp attraction is very
strong. This confirms the earlier COSY--11 findings~\cite{Winter}
but with higher statistics over much wider ranges of energy and
mass. The events that come from the non--$\phi$ region in the
2.65\,GeV data allow us to fix the parameter $a$ necessary to
describe the $K^-p/K^+p$ ratio within our final state interaction
model. The data are mainly sensitive to the magnitude $|a|$ and
the value obtained is not dissimilar to that required to describe
free \Kmp scattering.

The approach reproduces also our results at the other two energies
as well as the COSY--11 data. If allowance is made for the small
possible contribution to this ratio from $\phi$ events, of the
order of 20\%, the two--particle ratio is also well described for
all $K^+K^-$ mass intervals. In addition, we find a completely
analogous effect in the three-body $Kpp$ system, with the
$K^-pp/K^+pp$ ratio being strongly biased towards the lowest
values of the invariant mass. We show in Sect.~\ref{RatioKpp} that
these distributions are also predicted quantitatively by the same
\Kmp final state interaction model with the same value of the
parameter $a$.

The total cross sections for $\phi$ and non--$\phi$ production are
presented in Sect.~\ref{sigma_total}. The energy dependence of the
non--$\phi$ total cross section is strongly influenced by both the
$pp$ and \Kmp \emph{fsi} but, even after including these effects,
the lowest energy data are not well reproduced. This leads to the
suggestion that the effect might be connected with the two or
three points in the $K^+K^-$ spectrum at very low invariant
masses, which lie consistently above the simulations. Our
conclusions and views on the outlook are presented in
Sect.~\ref{Conclusions}.
%
%
\section{Experiment and data analysis}
\label{Experiment}%
The study of kaon pair production has been performed using data
taken with the ANKE magnetic spectrometer~\cite{Bar01}. This is
placed at an internal target station of COSY, which is the COoler
SYnchrotron of the Research Center J\"ulich~\cite{Mai97}. We have
already reported an initial analysis of the \reaction\ data, from
which values of the $pp\to pp\phi$ cross sections were
extracted~\cite{Har06}.

The measurements were performed at three proton kinetic energies
of $T_p=2.65\,$GeV, 2.70\,GeV and 2.83\,GeV, corresponding to
excess energies of %
$\varepsilon=51\,$MeV, 67\,MeV and 108\,MeV %
with respect to the $ppK^+K^-$ threshold. A dense hydrogen cluster
jet target~\cite{Mue99} provided areal densities of \mbox{$\sim
5\!\times\!10^{14}\,\textrm{cm}^{-2}$} and the average luminosity
during the experiment was typically $2.3 \times
10^{31}\textrm{cm}^{-2}\textrm{s}^{-1}$. The detection system of
the three--dipole magnetic spectrometer ANKE registers
simultaneously slow positively and negatively charged particles,
with fast positive particles being measured in the forward
system~\cite{Bar01,Dymov,Michael}.

Kaons produced in the $pp{\to}ppK^+K^-$ reaction at 2.65 and
2.70\,GeV have laboratory momenta between 0.2 and 1.0\,GeV/c and
this full range was covered by the detection system for both $K^+$
and $K^-$. For the 2.83\,GeV run, the $K^+$ were detected only up
to 0.6\,GeV/c, which reduced the geometrical acceptance. The
momentum range of the protons emerging from the reaction is
between 0.8 and 1.6\,GeV/c, for which the forward detector has
full acceptance.

The momenta of both the kaons and protons were determined from the
track information furnished by sets of multiwire proportional
chambers. The tracking efficiency for kaons fell smoothly from
98\% to 93\% as the momentum increased, whereas that for protons
was about 80\% over the whole momentum range. The efficiency map,
including its dependence on the particle momenta and position, was
generated for the track reconstruction and used in the analysis.
The data--taking efficiency was around 90\% and, taken in
combination with the tracking efficiency, resulted in a total
detector efficiency of about 60\%.

In order not to reduce the geometric acceptance unnecessarily,
only one forward--going proton was measured in coincidence with
the charged kaon pair. The $pp{\to}ppK^+K^-$ reaction was then
identified by requiring that the missing mass in the reaction was
consistent with that of the non--observed proton. As a first step,
positive kaons were selected through a procedure, described in
detail in Ref.~\cite{Bush}, that used the time of flight (TOF)
between START and STOP scintillation counters of a dedicated $K^+$
detection system. In the second stage, both the $K^{-}$ and
forward--going proton were identified from the time--of--flight
differences between the STOP counters in the negative and forward
detector systems with respect to the STOP counter in the positive
system that was hit by the $K^+$. By comparing this with the time
difference expected for a kaon pair based upon their measured
momenta, a clear separation is achieved, with good events lying in
an island where the two time differences are
consistent~\cite{Michael}. These two TOF selections, as well as
that for the $K^+$, were carried out within $\pm\,3\,\sigma$
bands.

\begin{figure}[h]
\centering
\includegraphics[width=6.8cm,clip]{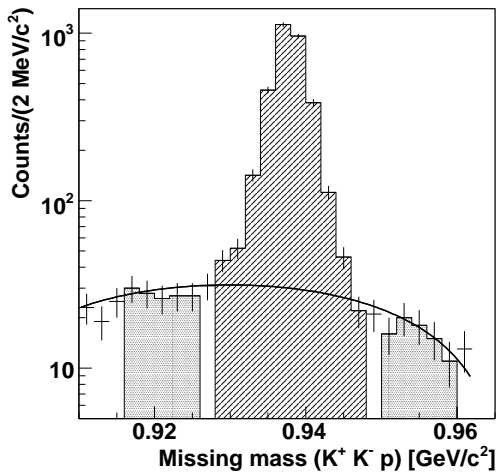}
\caption{Missing--mass distribution of $pp{\to}K^+K^-pX$ events at
$T_p$=2.65 GeV. The hatching indicates the range used for the
selection of protons. The side--band events scaled by the solid
line are indicated by the dotted shading.} \label{IMMM2.65}
\end{figure}

The spectrum of missing masses with respect to the $K^+K^-p$
system, shown in Fig.~\ref{IMMM2.65} for the 2.65\,GeV data, is
dominated by the proton peak. The corrected numbers of $K^+K^-pp$
events obtained in this way are around 3100 (at $T_p=2.65\,$GeV),
1300 (2.70\,GeV), and 650 (2.83\,GeV). The estimated background
inside the proton cut window is 5\%, 12\%, and 18\% at these three
energies, respectively. This has been subtracted using the
side--band events scaled by the solid line shown in
Fig.~\ref{IMMM2.65}. The uncertainty resulting from this procedure
is estimated to be less than 3\% for the total cross sections and
negligible compared to the statistical errors for the differential
distributions.

\begin{figure}[h!]
\centering
\includegraphics[width=6.8cm,clip]{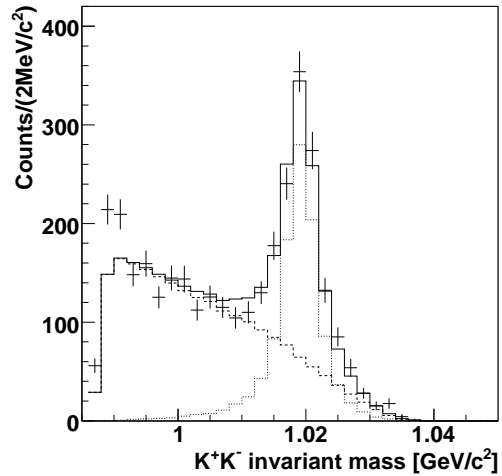}
\caption{The $K^+K^-$ invariant--mass distribution for the
\reaction\ reaction at 2.65\,GeV. The prominent $\phi$ peak sits
on a steadily falling background corresponding to the ANKE
acceptance for the four--particle final state. The simulation of
these two contributions and their sum are shown, respectively, by
the dotted, dashed, and solid histograms. \label{raw_spectrum_KK}}
\end{figure}

After the identification of the \reaction\ events, the data were
binned in intervals of the $K^+K^-$ invariant mass, with the
2.65\,GeV results being presented in Fig.~\ref{raw_spectrum_KK}.
This spectrum shows a very prominent $\phi$ peak positioned close
to the nominal mass of the meson. In addition there is a
non--$\phi$ contribution that decreases steadily from low to high
invariant masses. This behavior, which is mainly a reflection of
the ANKE acceptance for non--resonant $ppK^+K^-$ production, has
to be simulated in order to derive cross sections.

The acceptance has been determined through a Monte-Carlo
simulation based upon GEANT4~\cite{GEANT4}, where the geometrical
acceptance, resolution, detector efficiency, and kaon decay
probability were taken into account. In the initial step, the
$\phi$--production distributions previously published~\cite{Har06}
were taken as the basis of the simulation, whereas four--body
phase space was used for the non--$\phi$ component of
Fig.~\ref{raw_spectrum_KK}. There are seven degrees of freedom for
a four-body final state produced in an unpolarized reaction. Thus
for each beam energy, seven independent c.m.\ distributions were
generated (five angular distributions, the relative momentum of
final protons, and the $K^+K^-$ invariant mass). These were then
divided into two groups depending on the value of the $K^+K^-$
invariant mass, \emph{i.e.}, $\phi$--rich and $\phi$--poor
regions. All distributions were jointly fitted to the experimental
data and the relative contribution of $\phi$ and non--$\phi$
production evaluated for the determination of the acceptances.

In the $\phi$--poor region, most of the c.m.\ distributions are
fairly uniform, though the polar angle of the $K^+K^-$ system
relative to the beam axis in the overall c.m.\ system shows
deviations. Fitting this with the function $1+\alpha\cos^2\theta$
gives anisotropies of $\alpha=0.5\pm0.2$ and $\alpha=1.5\pm0.5$ at
2.65\,GeV and the two higher energies, respectively. The inclusion
of this factor increases the total acceptance for non--$\phi$
production by about 10\%. In addition, as will be seen in
Sects.~\ref{Ratio} and \ref{RatioKpp}, the $K^{\pm}p$ and
$K^{\pm}pp$ invariant masses deviate strongly from phase space.
These deviations have been taken into account in the simulations,
using iteratively the theoretical approach discussed in
Sect.~\ref{fsia}, in order to converge on the
acceptance--corrected distributions. The simulations of the $\phi$
and non--$\phi$ components of the uncorrected $K^+K^-$ spectrum at
2.65\,GeV are shown in Fig.~\ref{raw_spectrum_KK} together with
their sum.

The estimated total acceptance for kaon-pair production is 1.1\%
(2.65\,GeV), 0.7\% (2.70\,GeV) and 0.1\% (2.83\,GeV), and the
associated systematic uncertainties in the total cross section are
5\%, 7\% and 12\%, respectively. The shapes of the differential
cross sections, especially those for the $Kp$ and $Kpp$ invariant
masses, are very stable to this iterative procedure. As a
consequence, the uncertainty that this introduces in the
differential cross sections is negligible compared to the
statistical errors.

The luminosity required to evaluate absolute cross sections was
determined by measuring in parallel proton--proton elastic
scattering for laboratory angles between 5.0$^{\circ}$ and
8.5$^{\circ}$. For this purpose the momentum of a forward--going
proton was determined using the ANKE forward detector. After a
missing--mass analysis, the (small) background underneath the
proton peak was subtracted and the remaining events were corrected
for efficiency and acceptance.
%
%
\section{The $K^+K^-$ invariant mass distribution}
\label{KKbar}

\begin{figure}[h!]
\centering
\includegraphics[width=6.4cm,clip]{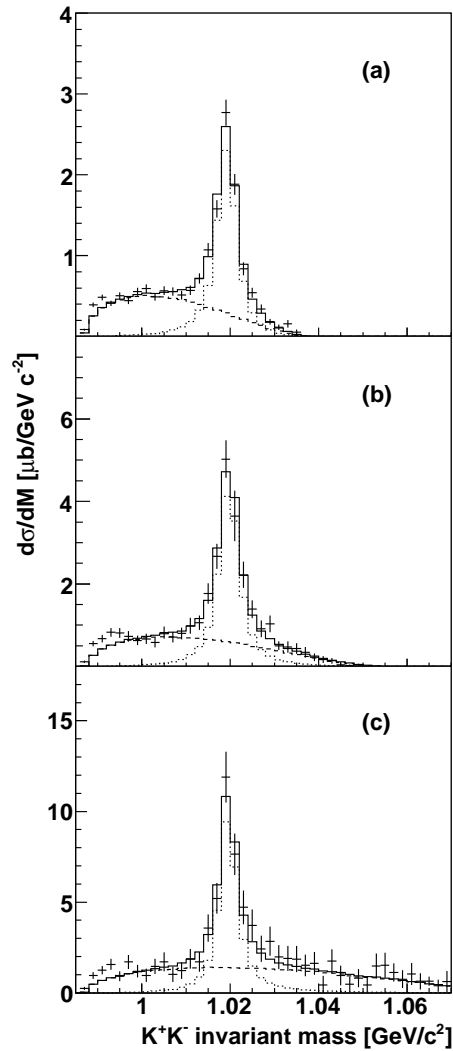}
\caption{Differential cross section for the \reaction\ reaction
with respect to the $K^+K^-$ invariant mass at (a) 2.65\,GeV, (b)
2.70\,GeV, and (c) 2.83\,GeV. The dotted histogram and the dashed
curve represent the simulations for the $\phi$ and non--$\phi$
contributions to the spectrum, with their sum being shown by the
solid histogram. For all three energies the simulations fail to
reproduce the lowest mass points.} \label{spectrum_KK}
\end{figure}

Figure~\ref{spectrum_KK} shows the \reaction\ differential cross
sections in terms of the $K^+K^-$ invariant masses as obtained at
our three energies. These have been corrected for acceptance,
deadtime, \emph{etc.}, and normalized on the basis of the
proton--proton elastic scattering data, as discussed in
Sect.~\ref{sigma_total}. The resulting non--$\phi$ contribution is
much flatter than that illustrated in Fig.~\ref{raw_spectrum_KK}.
In order to ensure a clean separation of the $\phi$ from the rest,
it is necessary to model the shapes as reliably as possible.

In our earlier work~\cite{Har06}, the non--$\phi$ contribution was
assumed to be described by a four--body phase space. The $\phi$
was taken to have a Breit--Wigner form with a constant width of
$\Gamma=4.26\,$MeV/c$^2$~\cite{PDG06}, which was then convoluted
with an experimental resolution of $\sigma=1\,$MeV/c$^2$. Only for
the $\phi$ component was the spectrum modified to include effects
arising from the final state interaction between the emerging
protons. These three assumptions have now to be reconsidered.

Since the $\phi$ is a $p$--wave resonance that decays strongly
into $K\bar{K}$, its width depends upon the $K^+K^-$ and
$K^0\bar{K}^0$ relative momenta. Due to the narrowness of the
state, the peak itself is little affected by this refinement but
it does suppress slightly the low mass tail of the resonance.
Although for low $K^+K^-$ effective masses there is no clear
evidence for any $pp$ \emph{fsi}, only minor effects are to be
expected there because of the large amount of energy that is then
available to excite the $pp$ system. Nevertheless, the \emph{fsi}
cannot be neglected at higher $K^+K^-$ masses, where its inclusion
distorts slightly the four--body phase space spectrum. Finally, as
will be seen in Sect.~\ref{Ratio}, it is clear that there is a
large final state interaction between the $K^-$ and one or both
protons and this also modifies the shape of the non--$\phi$
component, enriching the central part of the $K^+K^-$ spectrum. By
chance, all three effects go in the same direction. In the fits to
Fig.~\ref{spectrum_KK}, the fractions to be associated with $\phi$
production are diminished somewhat compared to those reported in
our previous work~\cite{Har06}.

It must be noted that the lowest mass points lie above the fitted
curves for all three data sets of Fig.~\ref{spectrum_KK}. These
deviations from the simulations are significant because, as seen
from the comparison of Figs.~\ref{raw_spectrum_KK} and
\ref{spectrum_KK}a, the ANKE acceptance is largest in this region.
Although ignored in the fitting process, they were taken into
account in the evaluation of the non--$\phi$ contribution to the
total cross section, contributing 4\% (2.65\,GeV), 6\%
(2.70\,GeV), and 10\% (2.83\,GeV).
%
%
\section{Final state interaction assumptions}
\label{fsia}

We wish to describe the \reaction\ data at all energies in terms
of final state interactions between the two protons and between
the $K^-$ and either one or both of the protons. There is no
reliable way of carrying out such a program without solving
multibody equations, which is well outside the scope of the
present paper. As explained in Appendix~\ref{app_a}, we make the
\emph{ad hoc} assumption that the overall enhancement factor is
the product of the enhancements in the $pp$ and two \Kmp systems,
all evaluated at the appropriate relative momenta $q$:
\begin{equation}
\label{assumption} F = F_{pp}(q_{pp})\times F_{Kp}(q_{Kp_1})
\times F_{Kp}(q_{Kp_2})\,.
\end{equation}
Similar approaches are to be found elsewhere in the literature as,
for example, in the description of the $pp\eta$ final state at low
energies~\cite{Bernard}.

The \Kmp enhancement factor is taken in the scattering length
approximation
\begin{equation}
\label{Kpfactor} F_{Kp}(q) =\frac{1}{1-iqa}\,,
\end{equation}
where some caution should be exercised in the interpretation of
the complex parameter $a$ as an effective scattering length. In
principle $a$ could have an energy dependence, which would
correspond to the introduction of an effective range term.
However, it is hard from our data to determine the values of any
extra free parameters.

\begin{figure}[hbt]
\centering \centerline{\epsfxsize=6.8cm {\epsfbox{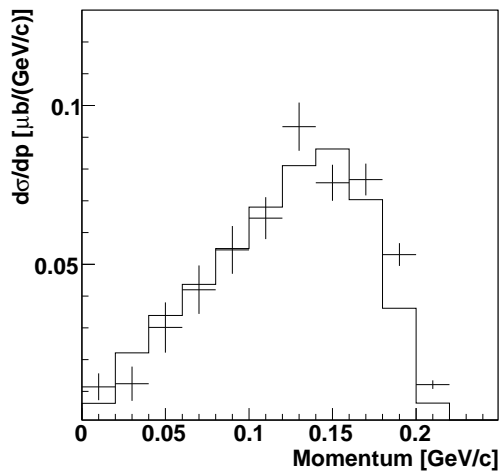}}}
\caption{Relative momentum of the final protons from the
\reaction\ reaction in the $pp$ rest frame. The experimental
points at 2.65\,GeV are compared to the simulation, using the
standard parameters, which is shown as a histogram.
\label{p_dist}}
\end{figure}

A proton--proton enhancement factor of the form
\begin{equation}
\label{ppfactor} |F_{pp}(q)|^2 = \frac{q^2+\beta^2}{q^2+\alpha^2}
\end{equation}
is assumed~\cite{GandW}, where the position of the $^{1\!}S_{0}$
virtual state is well fixed at $\alpha=0.1\,$fm$^{-1}$. The final
state interaction should not influence higher partial waves and,
in order to reduce its effects above about 10\,MeV, we take
$\beta=0.5\,$fm$^{-1}$.
The relative momentum of the two protons in the $pp$
rest frame, which is very sensitive to the form of the $pp$
\emph{fsi}, is then reproduced reasonably, as seen from
Fig.~\ref{p_dist}.

It is important to note that the form of neither the $pp$ nor the
\Kmp enhancement factor is valid for an excitation energy below
about 1\,MeV due to the explicit Coulomb force. However, these
regions represent only very small fractions of the allowed phase
space.

%
%
\section{The kaon--proton invariant--mass distribution}
\label{Ratio}

\begin{figure}[h!]
\includegraphics[width=6.4cm,clip]{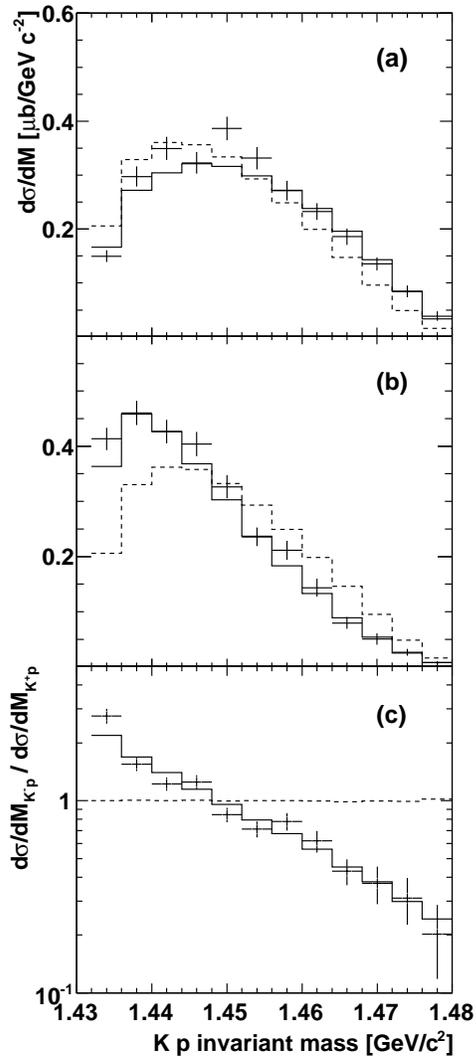}
\caption{Differential cross sections for the \reaction\ reaction
with respect to the invariant mass of (a) the $K^+p$ and (b) the
\Kmp systems. In panel (c) their ratio is shown on a logarithmic
scale. These 2.65\,GeV data have been selected to come from the
region $M(K^+K^-)< 1.01\,$GeV/c$^2$, where there is only a very
small $\phi$ contribution. The dashed histograms represent the
results of four--body phase--space simulations, where the only
distortion is that coming from the $pp$ \emph{fsi}. The solid
histograms include also the \Kmp \emph{fsi}.\label{265lt1010}}
\end{figure}

In data taken with the COSY--11 spectrometer at
$\varepsilon=28\,$MeV, \emph{i.e.}\ at an energy where there is
very little $\phi$ production, it was found that the \Kmp and
$K^+p$ invariant--mass $M(Kp)$ distributions were markedly
different~\cite{Winter}. The ratio of the acceptance--corrected
distributions
\[R_{Kp}=\frac{d\sigma/dM(K^-p)}{d\sigma/dM(K^+p)}\]
showed a very strong preference for low values of $M(Kp)$.

Figure~\ref{265lt1010} shows the $Kp$ spectra obtained at our
lowest energy. In order to consider a situation similar to that of
COSY--11, the data have been selected such that only events with
$K^+K^-$ invariant masses less than 1.01\,GeV/c$^2$ are retained.
The simulation that takes into account only the $pp$ \emph{fsi}
fails to describe the data. Furthermore, the ratio of the \Kmp to
$K^+p$ mass distributions is far from constant, as it would be if
there were only the $pp$ \emph{fsi}. The peaking of $R_{Kp}$ to
the lowest invariant masses confirms the COSY--11
observation~\cite{Winter} but with higher statistics and over a
wider range of $Kp$ masses.

\begin{figure}[h!]
\includegraphics[width=7.65cm,clip]{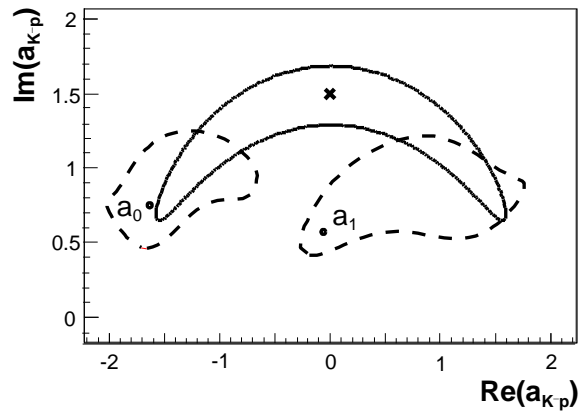}
\caption{Contour plot corresponding to the one--$\sigma$ level
uncertainty of the fits of the complex parameter $a$ of the
\emph{fsi} \textit{ansatz} to the data of Fig.~\ref{265lt1010}.
Our best fit is obtained with $a=1.5i\,$fm (marked with a cross).
Also shown are the best fit values (points) and corresponding
contours (dashed lines) for the isoscalar ($a_0$) and isovector
($a_1$) \Kmp scattering lengths derived from the study of free
$K^-$--nucleon scattering~\cite{Borasoy}. \label{contour}}
\end{figure}

When the \Kmp final--state--interaction factors are also
introduced into the four--body phase space simulation of the
non--$\phi$ contribution to the \reaction\ reaction, the
individual $d\sigma/dM(K^+p)$ and $d\sigma/dM(K^-p)$ distributions
of Fig.~\ref{265lt1010}, as well as their ratio $R_{Kp}$, are all
well described. The best fit to these data is achieved with an
effective \Kmp scattering length of $a= (0+1.5i)$\,fm and this
value of $a$ will be retained for all the subsequent simulations.
Nevertheless, the uncertainties in the real and imaginary parts
are quite large and strongly correlated such that $|a|$ is
determined much better than its phase, as illustrated by the
$\chi^2$ contour plot of Fig.~\ref{contour}. The free \Kmp $I=0$
and $I=1$ values from Ref.~\cite{Borasoy} are also shown there
along with their one--$\sigma$ limits. From this it is seen that
there is an overlap with either isospin scattering length. One has
also to bear in mind that our data represent averages over some
energy range and that a constant scattering length is necessarily
an oversimplification.

\begin{figure}[h!]
\includegraphics[width=5.95cm,clip]{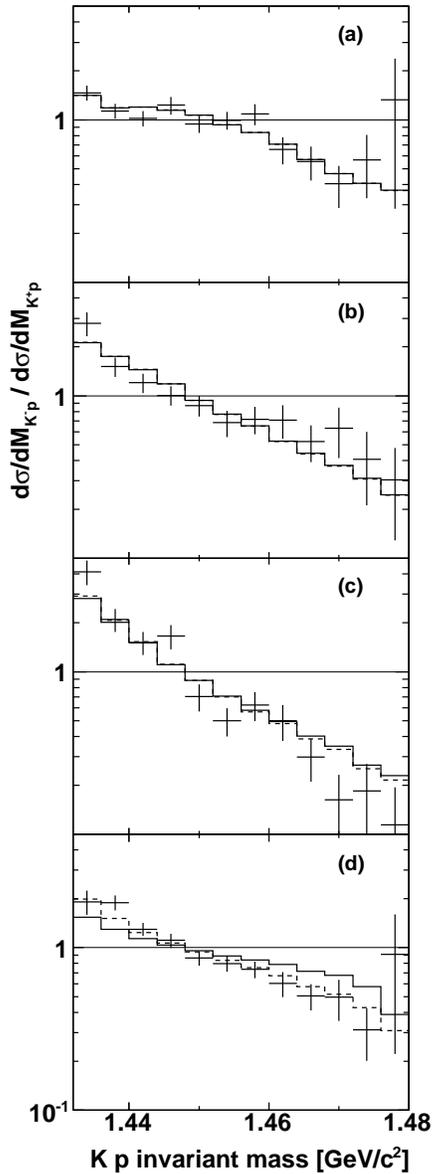}
\caption{Values of the $R_{Kp}$ ratio for the \reaction\ reaction
at 2.65\,GeV extracted for four different selections of the
$K^+K^-$ invariant mass, \emph{viz.}\ (a) $M(K^+K^-)<
0.995\,$GeV/c$^2$, (b) $0.995 < M(K^+K^-)< 1.003\,$GeV/c$^2$, (c)
$1.003 < M(K^+K^-)< 1.010\,$GeV/c$^2$, and (d) $1.010$GeV/c$^2 <
M(K^+K^-)$. The histograms are predictions of the model described
in the text, where the $K^-$ mesons from non--$\phi$ events are
subject to a \Kmp final state interaction. In the high mass
interval, the effects of assuming a 20\% probability of the $K^-$
from the $\phi$ decay being influenced by a \emph{fsi} is
considered and this leads to the dashed histograms.
\label{FourbinsMpK}}
\end{figure}

The situation for higher $K^+K^-$ masses is complicated by the
contribution from the $\phi$ meson. At our beam energies, the
average separation of the $K^-$ (produced in the $\phi$ decay)
from one of the final protons is about 7\,fm. This large distance
reduces considerably the effects of the \Kmp \emph{fsi}. We would
therefore expect that deviations of $R_{Kp}$ from unity should
arise primarily from the non-$\phi$ contributions. The modelling
of the data thus depends in the first instance upon the
determination of the fraction of $\phi$ production in the overall
\reaction\ spectra presented in Fig.~\ref{spectrum_KK}. However,
classical simulations suggest that more than 15\% of the $\phi$
decays lead to at least one \Kmp pair with a separation of less
than 1\,fm, for which the \emph{fsi} is not negligible.

\begin{figure}[htb]
\includegraphics[width=5.95cm,clip]{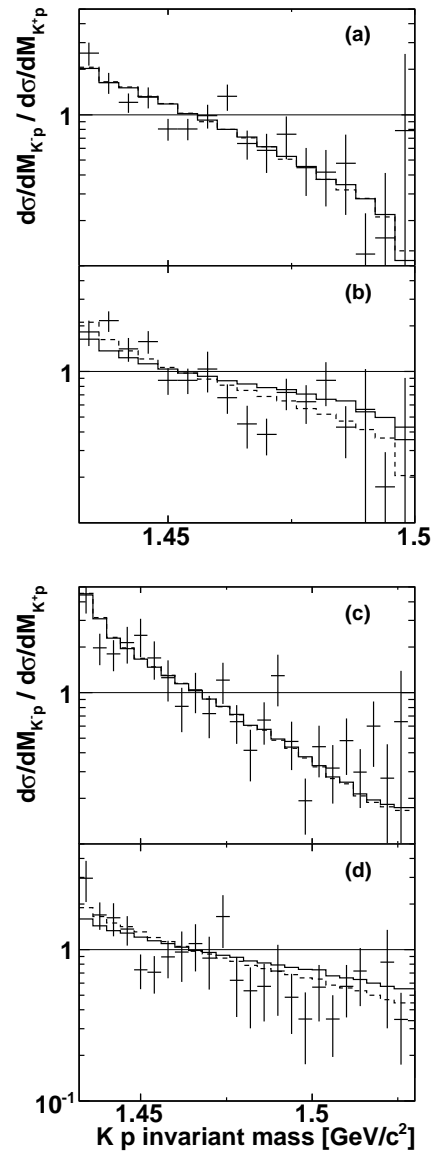}
\caption{Values of the $R_{Kp}$ ratio for (a) the $\phi$--poor
region and (b) the $\phi$--rich  at 2.70\,GeV with (c) and (d)
showing respectively the analogous results at 2.83\,GeV. The
predictions obtained using the standard parameters are shown with
no \emph{fsi} effect for the $K^-$ from the decay of the $\phi$
(solid curve) and 20\% (dashed). \label{ratio270283}}
\end{figure}

To investigate this point further, the 2.65\,GeV data have been
divided into four regions of $K^+K^-$ invariant mass, \emph{viz.}\
$M(K^+K^-)< 0.995\,$GeV/c$^2$, $0.995 < M(K^+K^-)<
1.003\,$GeV/c$^2$, $1.003 < M(K^+K^-)< 1.010\,$GeV/c$^2$, and
$1.010$GeV/c$^2 < M(K^+K^-)$. The first two intervals contain
almost no $\phi$ events, whereas the last corresponds mainly to
$\phi$ production. The corresponding $R_{Kp}$ distributions are
shown in Fig.~\ref{FourbinsMpK}, where they are compared to the
same simulation as that shown in Fig.~\ref{265lt1010}. The
non-$\phi$ region is very well described but, for high $K^+K^-$
invariant masses, the asymmetry seems to be diluted too much by
kaons from the $\phi$ decay. The agreement is improved by assuming
that 20\% of the $\phi$ events are influenced by the \Kmp
\emph{fsi}. The same effect could be achieved by artificially
reducing the fraction of $\phi$ mesons in the peak of
Fig.~\ref{spectrum_KK}a by 10\%, but this would be hard to
justify.

If the strong \Kmp final state interaction is responsible for the
distortion in the $K^-p/K^+p$ ratio seen in
Fig.~\ref{FourbinsMpK}, similar effects should be seen for the
other excess energies. For our $\varepsilon =66.6$ and 108\,MeV
results shown in Fig.~\ref{ratio270283}, the statistics are lower
and we merely divide each data set into two, comprising
$\phi$--rich and $\phi$--poor regions. Without any change in the
value of the \Kmp effective scattering length, all the
distributions are well described, especially if it is assumed that
20\% of the $\phi$ events lead to \Kmp \emph{fsi} effects.

\begin{figure}[h]
\centering
\includegraphics[width=5.95cm,clip]{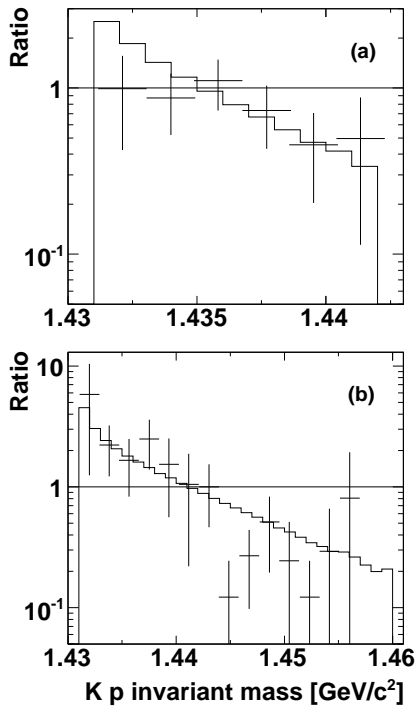}
\caption{The $R_{Kp}$ ratio for the COSY--11 \reaction\ data at
(a) $\varepsilon=10\,$MeV (upper panel) and (b)
$\varepsilon=28\,$MeV. The small contribution from the low mass
tail of the $\phi$ has been neglected in the 28\,MeV simulation.
Since the 10\,MeV results were arbitrarily scaled in the
publication~\cite{Winter}, they have here been divided by a factor
of three before being compared to the predictions of the
\emph{fsi} model, using the same values of the parameters as for
the ANKE data. } \label{COSY11}
\end{figure}

Data on the $K^-p/K^+p$ ratio were not published by the DISTO
collaboration~\cite{Balestra}. However, the same good description
of $R_{Kp}$ is found for the COSY--11 results of Fig.~\ref{COSY11}
where, in view of the low excess energies, there is no ambiguity
regarding the $\phi$ contribution. Given that the systematic
uncertainties in this experiment are independent of those at ANKE,
the quantitative agreement shown by the 28\,MeV data is
convincing. Unfortunately, in the publication~\cite{Winter} the
values of the ratio at 10\,MeV were scaled upwards to make them
lie closer to those taken at 28\,MeV. Since the numbers of \Kmp
and $K^+p$ events are the same, the weighted average of the ratio
must be unity and this allows one to make an informed guess
regarding the value of the scaling that has to be reintroduced. No
account has been taken of the COSY--11 invariant--mass resolution,
which will be of greater significance for the 10\,MeV data.

%
%
\section{The Kaon--proton--proton invariant--mass distribution}
\label{RatioKpp}

\begin{figure*}[htb]
\centering
\includegraphics[width=0.9\textwidth,clip]{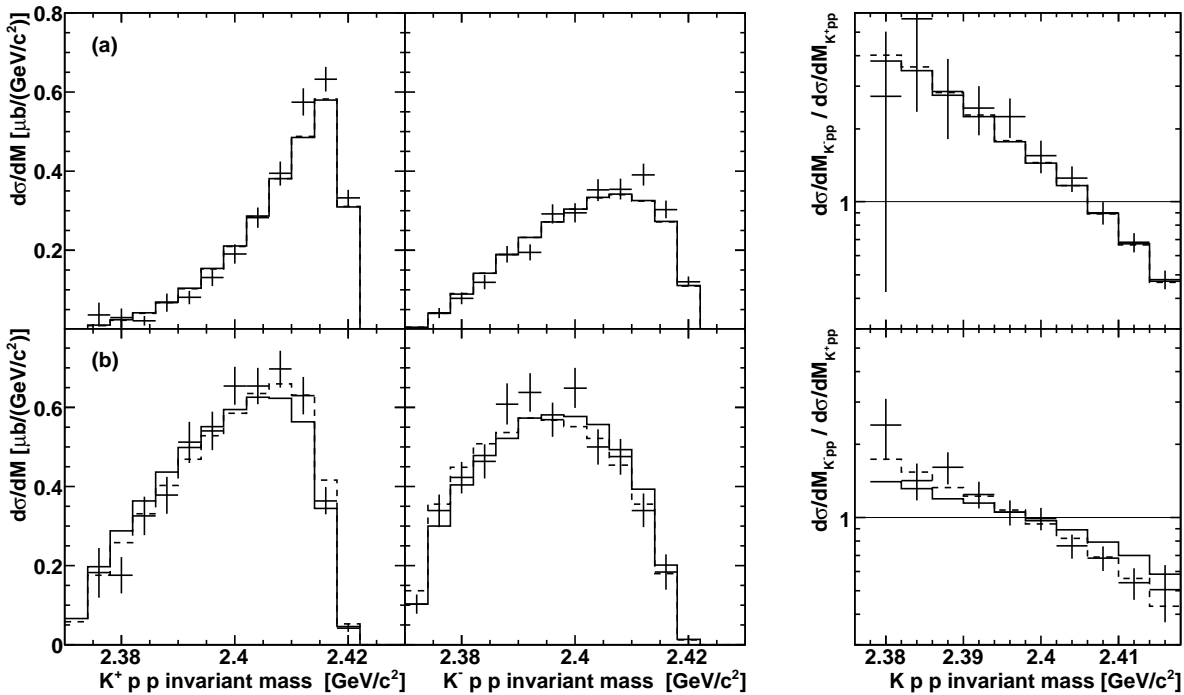}
\caption{Differential cross sections for the \reaction\ reaction
at 2.65\,GeV with respect to the invariant masses of the $K^+pp$
and the $K^-pp$ systems and their ratio. The data are divided (a)
into a $\phi$--poor region with $M(K^+K^-) <
1.01\,\textrm{GeV/c}^2$ (upper panels) and (b) $\phi$--rich with
$M(K^+K^-)> 1.01\,\textrm{GeV/c}^2$ (lower panels). The
predictions obtained using the standard parameters are shown with
no \emph{fsi} effect for the $K^-$ from the decay of the $\phi$
(solid curve) and 20\% (dashed). The simulations that describe the
two--particle ratios also represent well these three--body ratios.
} \label{Kpp}
\end{figure*}

We have argued in Sect.~\ref{fsia} and Appendix~\ref{app_a} that
it is natural to take the final state enhancement to be the
product of the individual factors corresponding to the $pp$ and
two \Kmp interactions, which allows the kaon to be attracted
simultaneously to both protons. This in turn suggests that there
might also be an enhancement in the three--body system at low
$K^-pp$ invariant masses. This is indeed the case. The size of the
effect at 2.65\,GeV is illustrated in Fig.~\ref{Kpp} for the
regions with $M(K^+K^-) \gtrless 1.01\,\textrm{GeV/c}^2$.

\begin{figure}[htb]
\centering
\includegraphics[width=5.78cm,clip]{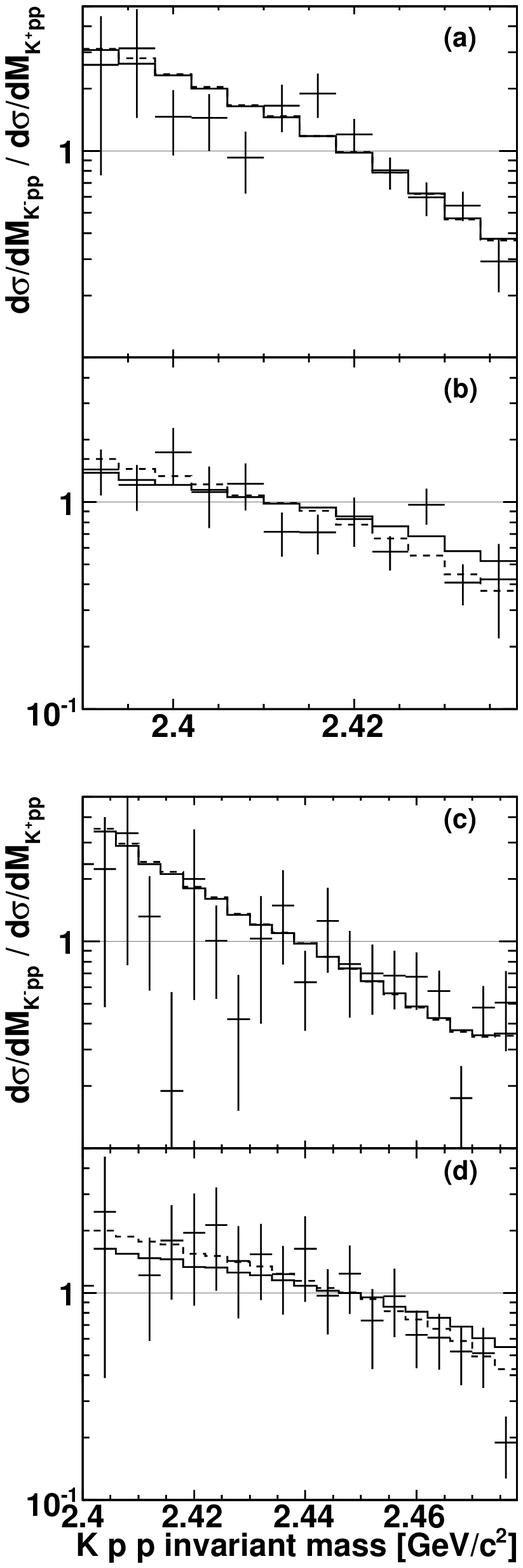}
\caption{The ratio of the differential cross sections of the
\reaction\ reaction with respect to the $K^-pp$ to $K^+pp$
invariant masses for (a) the $\phi$--poor region and (b) the
$\phi$--rich  at 2.70\,GeV with (c) and (d) showing respectively
the analogous results at 2.83\,GeV. The predictions obtained using
the standard parameters are shown with no \emph{fsi} effect for
the $K^-$ from the decay of the $\phi$ (solid curve) and 20\%
(dashed). The simulations that describe the two--particle ratios
also represent well these three--body ratios. \label{Kpp2}}
\end{figure}

There is a striking preference for low $Kpp$ masses in both
$K^+K^-$ intervals and this behavior is reproduced quantitatively
by our simulation, which uses the same \Kmp effective scattering
length $a$ that was chosen to describe the two--body $R_{Kp}$
ratio. It provides some empirical justification for our choice of
the final--state--interaction enhancement factor as the product of
the individual two--body enhancements. This favors configurations
where the $K^-$ is attracted simultaneously to both protons. It
must, however, be stressed that the strong mass dependence which
is apparent in Fig.~\ref{Kpp} does not necessarily imply that the
$K^-$ forms a bound state with the two
protons~\cite{Agnello,Yamazaki,Gal}.

Though the error bars are considerably larger at the two higher
energies, as seen in Fig.~\ref{Kpp2}, the simulation is consistent
with the results. In all these three--particle mass distributions
the introduction of a small amount of \emph{fsi} for the products
of the $\phi$ decay seems to be of less importance than for the
$Kp$ spectra.
%
%
\section{Total cross sections}
\label{sigma_total}

The total cross section results for both $\phi$ and non--$\phi$
production are given in Table~\ref{table1} for the three energies.
The systematic errors given here arise from the quadratic sum of
the uncertainties in the background subtraction, acceptance
correction, tracking efficiency correction for kaons (3\%) and the
analysis of the $pp$--elastic data used for the normalization.
Unfortunately, the $pp$ database is very limited in our region of
angle and energy. This is recognized in the recent SP07 update to
the SAID program where it is stated that \emph{our solution should
be considered at best qualitative between 2.5 and
3\,GeV}~\cite{SAID-update}. The raw SP07 solution generally
overestimates the values of those $pp$ differential cross sections
that are available in this range~\cite{ppdata}. A normalization
option has to be chosen to ensure agreement with the $pp$
experimental data and this typically leads to reductions of about
15\% at our two lower energies but almost 20\% at 2.83\,GeV. To
eliminate any ambiguities, the values that we have assumed for the
$pp$ elastic cross section, integrated over the solid angle
defined by $5.0^{\circ}< \theta_{\text{lab}} < 8.5^{\circ}$, are
presented in Table~\ref{table1} for our three energies.

\begin{table}[hbt]
\caption{\label{table1}%
Total cross sections for the $pp\to
pp\{K^+K^-\}_{\text{non-}\phi}$ and $pp{\to}pp \phi$ reactions at
three beam energies $T_p$ and corresponding excess energies
$\varepsilon_{KK}$ with respect to the $ppK^+K^-$ threshold. The
excess energy with respect to the $\phi$ threshold is given by
$\varepsilon_{\phi}=\varepsilon_{KK}-32.1\,$MeV. The $\phi$ cross
section has been corrected for the 49.1\% branching
ratio~\cite{PDG06}. For both channels the first error is
statistical and the second systematic. The values that we have
taken for $\sigma_{pp}$, the integral of the proton--proton
elastic cross section over the solid angle range $5.0^{\circ} <
\theta_{\text{lab}} < 8.5^{\circ}$, are also given. The associated
$\approx\pm 6\%$ uncertainty has not, however, been compounded
with the other errors.}
\begin{ruledtabular}
\begin{tabular}{c|c|c|c|c}
$T_p$&$\varepsilon_{K^+K^-}$&$\sigma_{pp}$&
$\sigma_{\text{non-}\phi}$(tot)&$\sigma_{\phi}$(tot)\\
{}[GeV]&[MeV]&[mb]&[nb]&[nb] \\ \hline
2.65&\phantom{1}51 &$5.52$&$16\pm1\pm1$  & $33\pm2\pm4$\\
2.70&\phantom{1}67 &$5.44$&$30\pm2\pm3$  & $64\pm4\pm10$\\
2.83&108           &$5.06$&$98\pm8\pm15$ & $133\pm12\pm27$ \\
\end{tabular}
\end{ruledtabular}
\end{table}

The value that we obtain for the luminosity at 2.65\,GeV on the
basis of the $pp$ cross section given in the table could be
checked using a novel technique~\cite{Stein}. This depends upon
the energy loss of the coasting proton beam caused by its multiple
traversals through the very thin target. The resultant frequency
change can be measured with high accuracy using the spectrum of
the Schottky noise. Combined with measurements of the beam
current, this yielded luminosities with an expected precision of
about $\pm6\%$. The two methods give consistent results within
this uncertainty and, although the frequency--change technique
could not be applied as reliably at the higher energies, it is
expected that the overall luminosity uncertainty is on about this
level.

\begin{figure}[h]
\centering \centerline{\epsfxsize=1.1\columnwidth
{\epsfbox{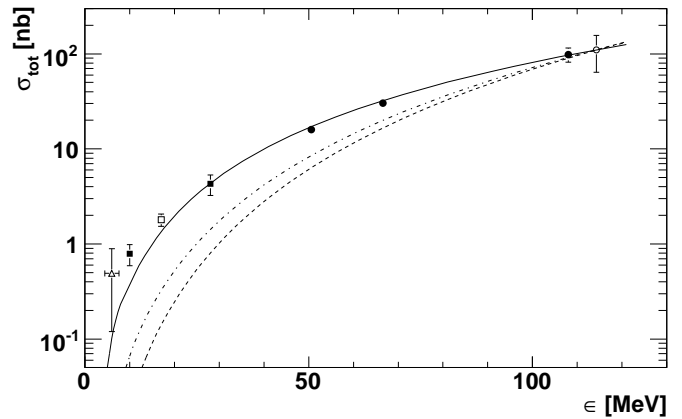}}} \caption{Energy dependence of the
non-$\phi$ contribution to the \reaction\ total cross section. In
addition to our three points (closed circles), there is a high
energy point from DISTO (open circle)~\cite{Balestra} and four
points close to threshold from COSY--11 (closed
squares)~\cite{Winter}, (open triangle)~\cite{Wolke}, and (open
square)~\cite{Quentmeier}. The dashed curve represents the energy
dependence from four--body phase space. The dot-dashed includes
the effects of the $pp$ final state interaction whereas the solid
curve contains in addition distortion from the \Kmp \emph{fsi}, as
described in the text.} \label{TotalXsec}
\end{figure}

Our total cross section results for non--$\phi$ production are
plotted in Fig.~\ref{TotalXsec} along with results taken from
DISTO~\cite{Balestra} and COSY--11~\cite{Winter,Wolke,Quentmeier}.
It is seen that four--body phase space cannot describe
simultaneously the \reaction\ data at high and low excess
energies. The situation is improved only a little if account is
taken of the strong attraction between the two protons in the
$^{1\!}S_{0}$ state but a greater improvement is achieved through
the introduction of the \emph{fsi} in the \Kmp system.
Nevertheless, the COSY--11 data at excess energies of 20\,MeV or
less still seem to be underestimated. This might be connected to
the incomplete description of the $K^+K^-$ spectra of
Fig.~\ref{spectrum_KK} at very low invariant masses. However, it
must be stressed that in the evaluation of the COSY--11 acceptance
no account was taken of any \Kmp \emph{fsi}.

%
%
\section{Conclusions and outlook}
\label{Conclusions}

We have investigated the \reaction\ reaction at three energies. In
those parts of the $K^+K^-$ mass spectra where $\phi$ production
is small, we find a marked difference between the \Kmp and $K^+p$
invariant--mass distributions. There is a strong peaking in the
$K^-p/K^+p$ ratio towards the $Kp$ threshold for both the COSY--11
and our data. This clearly indicates that the \reaction\ reaction
cannot be dominated by the undistorted production of a
\emph{single} scalar resonance $a_0$ or $f_0$. Either a mixture of
the two is produced or one of the kaons from the $a_0/f_0$ decay
interacts with a final proton. As we have shown, these data can be
explained quantitatively through the introduction of a simple \Kmp
final state interaction.

In kinematic regions where $\phi$ production is significant, the
$K^-\!p/K^+p$ asymmetry is reduced because the meson may travel
some distance before decaying. Our data in such conditions are
well fit by assuming that about 20\% of the $K^-$ from the $\phi$
decay interact with one or both final protons. The uncertainty in
this is, however, quite large because of the limited precision
with which the $\phi$ contribution to the $K^+K^-$ mass spectrum
can be isolated.

There is no universal technique for modelling the \emph{fsi} when
more than one pair of particles is involved. The product
\emph{ansatz} has the benefit that the same simulation with the
same parameters reproduces also the strong peaking in the ratio of
the $K^-pp/K^+pp$ mass distributions.

We find an effective \Kmp scattering length that is of the same
order of magnitude as that needed to describe the free scattering
data~\cite{Borasoy}. However, the error bars are large and our
data are primarily sensitive to $|a|$ rather than its phase. It is
important to stress that we do not know the relative strengths for
the production of isospin $I=0$ and $I=1$ \Kmp pairs in the
\reaction\ reaction and hence the relative weights of $a_0$ and
$a_1$. Some information on this might come from looking at the
data on the $pp\to K^+p\Sigma^0(1385)$ and $pp\to
K^+p\Lambda(1405)$ reactions taken at 2.83\,GeV~\cite{Iza}.
However, even this would not give access to the phase between the
$I=0$ and $I=1$ amplitudes. It must also be stressed that we have
neglected any energy dependence of the parameter $a$. Another
important \emph{caveat} is that we have not considered any
possible interaction of the $K^+$ with the protons. Although there
is no evidence for any strong effect of this nature in the $pp\to
K^+\Lambda p$ reaction~\cite{TOF}, since the $K^+p$ interaction is
repulsive, its neglect might be interpreted as extra attraction in
the \Kmp system.

We have clearly shown the importance of the \Kmp interaction,
which is possibly related to the $\Lambda(1405)$ or the
$\Sigma^0(1385)$, but we must now ask whether there is any
evidence for effects that might be connected with the production
of the $a_0/f_0$ resonances. The simulation of the energy
dependence of the total cross section for the production of the
non--$\phi$ component in the \reaction\ reaction underestimates
the COSY--11 data very near threshold. Even more intriguing, and
possibly connected, are the lowest mass points in the $K^+K^-$
spectra. At all three energies shown in Fig.~\ref{spectrum_KK},
these points lie much higher than the simulations and a similar
behavior is visible in the DISTO results~\cite{Balestra} as well
as in our data on the $pn\to dK^+K^-$ reaction~\cite{Mae06}. The
mass scale of the variation is not that of the widths of the
scalar resonances, which are quite large. It is tempting to
suggest that this structure might be due to the opening of the
$K^0\bar{K}^0$ channel at a mass of $\approx 8\,$MeV/c$^2$, which
induces some cusp structure that changes the energy dependence of
the total cross section near threshold~\cite{Winter,Walter}. This
would require a very strong $K^+K^-\rightleftharpoons
K^0\bar{K}^0$ channel coupling, which might be driven by the
$a_0/f_0$ resonances. Thus, although the \reaction\ reaction may
not be ideal for investigating the properties of scalar states,
their indirect effects might still be crucial.

Given that the $K^-$ is strongly attracted to protons, it is
natural to expect that similar \emph{fsi} effects should exist for
other reactions. The $\bar{K}^0d/K^+d$ ratio in the $pp\to
K^+\bar{K}^0d$ reaction does indeed show a very strong enhancement
near the $Kd$ threshold~\cite{Vera}. The interpretation is,
however, slightly more complex because spin--parity constraints
means that there must be at least one $p$--wave in the final
state, even at low energies. Unfortunately, the MOMO data on $pd
\to K^+K^-\,^3$He~\cite{MOMO} cannot be used to isolate any
$K^-\,^3$He interaction because the charges of the kaons were not
identified and only averaged spectra could be
studied~\cite{Markus}.

There is currently an overall uncertainty in our values of the
cross sections because of the limitation in the proton--proton
elastic scattering database in our region of energy and angle,
though we hope that this will be alleviated by future
measurements~\cite{Stein}. With our new estimates for the
luminosity, the $\phi$--production cross sections are reduced
compared to those of our first analysis~\cite{Har06}, which relied
upon the non--renormalized 2004 SAID solution~\cite{SAID-update}.
Furthermore, the better separation of the $\phi$ and non--$\phi$
events goes by chance in the same direction. These two effects
lower the ratio of $\phi$ to $\omega$ production in proton--proton
collisions near threshold so that it is now only about a factor of
six above the Okubo--Zweig--Iizuka limit~\cite{OkuboZweigIizuka}.
%
%
\begin{acknowledgments}

We would like to thank the COSY machine crew for their support as
well as that of other members of the ANKE Collaboration.
Discussions with J.~Haidenbauer, C.~Hanhart and A.~Sibirtsev were
very helpful and those with I.~Strakovsksy provided important
clarifications on the uncertainties in the SAID $pp$ procedure
above 2.5\,GeV. R.~Ni{\ss}ler and U.--G.~Mei{\ss}ner kindly provided
us with the predictions of Ref.~\cite{Borasoy} in numerical form.
Correspondence with R.~Bertini and P.~Winter is gratefully
acknowledged. This work has been supported in part by the BMBF,
DFG, Russian Academy of Sciences, and COSY FFE.
\end{acknowledgments}
%
%
\begin{appendix}
\section{Final state interactions}
\label{app_a} When two particles interact strongly in the final
state, the resulting matrix element involves an average of a
production operator with the relative wave function
$\psi_{\vec{q}}(\vec{r})$ of the strongly interacting pair. If the
interaction is of very short range, the wave function may be
evaluated at the origin to leave an enhancement factor
\begin{equation}
\label{A1} F_2(q)\propto \psi_{\vec{q}}(0)\propto
\frac{1}{D(q)}\,,
\end{equation}
where $\vec{q}$ is the relative momentum in the pair and $D(q)$ is
the $S$--wave Jost function~\cite{GandW}. In the commonly used
scattering length approximation, one retains only the linear term
in $q$, in which case $F(q)=1/(1-iqa)$, where $a$ is the
scattering length of the interacting pair.

On the other hand, if two or more pairs of particles interact in
the final state, there is no reliable prescription to evaluate an
analogous enhancement factor since a three--body equation then
needs to be solved. If we denote the interacting pairs as 12 and
13, the corresponding wave function will be
$\Psi_{\vec{q}_{12},\vec{q}_{13}}(\vec{r}_{12},\vec{r}_{13})$. We
now make the \emph{ad hoc} assumption that this wave function
factorizes in the form
\begin{equation}
\label{A2}\Psi_{\vec{q}_{12},\vec{q}_{13}}(\vec{r}_{12},\vec{r}_{13})
= \psi_{\vec{q}_{12}}(\vec{r}_{12})\times
\psi_{\vec{q}_{13}}(\vec{r}_{13})\,.
\end{equation}
In this case, the three--body enhancement factor is simply the
product of the two--body factors, evaluated at the appropriate
relative momenta:
\begin{eqnarray}
\nonumber
F_3(q_{12},q_{13}) &=& F_2(q_{12})\times F_2(q_{13})\\
&\approx& \frac{1}{(1-iq_{12}a_{12})(1-iq_{13}a_{13})}\,.
\label{A3}
\end{eqnarray}

It must be stressed that Eq.~(\ref{A3}) is merely an \emph{ansatz}
to try to understand our ensemble of data and, even if it provides
a satisfactory description of these, this does not mean that it is
applicable more generally. However, it has also been used to treat
the $pp\to pp\eta$ reaction near threshold, where all the final
pairs of particles interact strongly~\cite{Bernard}.

The approach does retain the necessary pole structure when
$q_{12}=-i/a_{12}$ and similarly for $q_{13}$. Thus particle-1 can
interact simultaneously with both 2 and 3. Furthermore, to lowest
order in the momenta, the \emph{ansatz} corresponds to the
scattering length of particle--1 from a composite 2+3 with the
desired combined scattering length of $a=a_{12}+a_{13}$.

Although the $S$--wave $pp$ interaction is well known, the \Kmp
interaction is far more complex because of the channel couplings
to $\Sigma\pi$ and also because there are two isospins $I=0$ and
$I=1$. In a recent study, within a chiral SU(3) unitary
approach~\cite{Borasoy}, \Kmp data leading to a variety of
channels have been fit and values of the (complex) \Kmp scattering
lengths deduced for $I=0$ and $I=1$. Their \emph{full model}
resulted in
\begin{eqnarray}
\nonumber
a_0&=&(-1.64+i0.75)\,\textrm{fm}\,,\\
a_1&=&(-0.06+i0.57)\,\textrm{fm}\,. \label{A4}
\end{eqnarray}
These values describe the rapid drop in the \Kmp elastic
amplitudes from threshold.

If isospin breaking induced by the significant mass differences is
neglected, the \Kmp scattering length $a_{K^-p}=(a_0+a_1)/2$.
However, this is not necessarily the parameter that is relevant
for the analysis of the $pp\to pK^+\{K^-p\}$ because it is not
clear whether the reaction mechanism preferentially excites $I=0$
or $I=1$ $\{K^-\!p\}$ states or some combination thereof. The
values in Eq.~(\ref{A4}) should therefore be considered merely as
order--of--magnitude estimates of what might be expected when
making fits to our data.
\end{appendix}
%
%

\end{document}